\documentclass[aps,twocolumn,prb,floatfix, superscriptaddress]{revtex4-2}

\usepackage[T1]{fontenc}
\usepackage{mathptmx} 
\usepackage[english]{babel}
\usepackage[utf8]{inputenc}
\usepackage{polski}
\usepackage{indentfirst}
\usepackage{amsthm}
\usepackage{amsmath}

\usepackage{lmodern} 
\usepackage{graphicx}
\usepackage{geometry}
\usepackage[caption=false]{subfig}
\usepackage{hyperref}
\usepackage{relsize}
\usepackage{textgreek}
\usepackage{gensymb}

\usepackage[utopia]{mathdesign} 

\graphicspath{ {images/} }

\usepackage[colorinlistoftodos]{todonotes}

\usepackage[colorinlistoftodos]{todonotes}

\usepackage[colorinlistoftodos]{todonotes}

\usepackage[colorinlistoftodos]{todonotes}

\usepackage[colorinlistoftodos]{todonotes}

\usepackage{babel}

\newcommand\I{\mathrm{i}}
\newcommand\E{\mathrm{e}}

\newcommand\tsub[1]{_{\text{#1}}}

\newcommand\vect[1]{\boldsymbol{#1}}
\newcommand\mat[1]{\mathsf{#1}}
\newcommand\abs[1]{\lvert #1\rvert}

\newcommand\smallfactor[1]{\underline{#1}}

\DeclareMathOperator{\RE}{Re}

\newgeometry{tmargin=1.5cm, bmargin=1.5cm, lmargin=1.5cm, rmargin=1.5cm}

\begin{document}

\preprint{APS/123-QED}
\title{Resonant subwavelength control of the phase of spin waves reflected from a ferromagnetic film edge}

\author{Krzysztof Sobucki}
\email{krzsob@st.amu.edu.pl}
\affiliation{Faculty of Physics, Adam Mickiewicz University, Uniwersytetu Poznańskiego 2, 61-614 Poznań, Poland}

\author{Wojciech Śmigaj}
\affiliation{Met Office, FitzRoy Rd, Exeter, EX1 3PB, UK}

\author{Justyna Rychły}
\affiliation{Institute of Molecular Physics, Polish Academy of Sciences, Mariana Smoluchowskiego 17, 60-179 Poznań, Poland}

\author{Maciej Krawczyk}
\affiliation{Faculty of Physics, Adam Mickiewicz University, Uniwersytetu Poznańskiego 2, 61-614 Poznań, Poland}

\author{Paweł Gruszecki}
\email{gruszecki@amu.edu.pl}
\affiliation{Faculty of Physics, Adam Mickiewicz University, Uniwersytetu Poznańskiego 2, 61-614 Poznań, Poland}
\affiliation{Institute of Molecular Physics, Polish Academy of Sciences, Mariana Smoluchowskiego 17, 60-179 Poznań, Poland}

\begin{abstract}

    Using frequency-domain finite element calculations cross-checked with micromagnetic simulations, we demonstrate that the phase of spin waves reflected from an interface between a permalloy film and a bilayer can be controlled by changing dimensions of the bilayer. Treating the bilayer formed by the permalloy film and a ferromagnetic stripe as a segment of a multi-mode waveguide, we show that spin-wave Fabry-Perot resonances of one of its modes are responsible for the high sensitivity of the phase of reflected waves to stripe width and the stripe-film separation. Thus, the system is a unique realization of a fully magnonic Gires–Tournois interferometer based on a two-modes resonator, which can be treated as a magnonic counterpart of a metasurface, since it enables manipulation of the phase of spin waves at subwavelength distances. Knowledge gained from these calculations might be used to design magnonic devices such as flat lenses or magnetic particle detectors.

   \end{abstract}

\date{\today}
\maketitle

\section{Introduction}

The recent years have been marked by a rapidly growing demand for interconnected mobile devices. This emerging ecosystem of connected devices, preferably communicating wirelessly, is referred to as the Internet of Things. There are estimations that within the next few years, the number of WiFi-enabled devices will be at least four times larger than the total population of the world \cite{ng2017internet}. One of the essential components of the Internet of Things are small and energetically efficient devices processing signals converted from and then back to microwaves. In this field, the application of spin waves (SWs), which are collective disturbances of magnetization propagating in the same frequence range as microwaves and thus able to couple to them, opens up a new opportunity to increase the efficiency of these devices. Compared to existing microwave devices, SW components offer prospects for increased miniaturization (SWs have wavelengths 3--5 orders of magnitude shorter than microwaves of the same frequency), easy external control of SW signals, reprogrammability, and small energy demands due to lack of Joule heating related with SWs propagation \cite{krawczyk14rev, chumak2015rev,chumak2019fundamentals}.

In order to use any kind of waves as an information carrier, efficient methods of their excitation and control over their amplitude and phase must be developed. 
In modern photonics, a breakthrough in the control of these properties of reflected or transmitted waves at subwavelength distances has recently been achieved through the use of arrays of nanostructured antennas absorbing and reemitting modified electromagnetic waves \cite{Yu2014,yu2011light}. These arrays, so-called metasurfaces, are used to obtain anomalous refraction of incident waves or to design flat, ultra-narrow lenses able to focus waves. Moreover, such nanostructured antennas can serve as color filters that can be used to produce silicon- or metallic-based color pixels for printing purposes as a replacement of chemical dyes \cite{Kumer2012,vashistha2017colorFilters}.

There are several reports on  the coupling of small magnetic elements with an uniform film.
Kruglyak et al. have shown that a narrow ferromagnetic element placed on top of a magnetic waveguide can be used to emit SWs, to control the phase of SWs passing below the resonator, and under some conditions even to absorb the energy of propagating SWs \cite{kruglyak2017graded,au2012phaseshifter,au2012transducer}. Yu at al. have demonstrated chiral excitation of SWs in a thin film through its dipolar coupling with a single nanowire or a grating of nanowires  placed in a spatially uniform microwave-frequency magnetic field \cite{yu2019wire,yu2019gratting}.
Subsequently, the existence of Fano resonances and their influence on the amplitude and phase of transmitted waves in a single-mode waveguide has been further studied by Al et al.~\cite{al2008evidence}, and Zhang et al.\ have demonstrated the application of a single dynamically tunable resonator in zero bias field placed on top of a waveguide to tune the phase of the transmitted SWs \cite{zhang2019resonator}. A grating coupler made up of an array of resonators has been used to excite short-wavelength SWs \cite{yu2013gratingCoupler,yu2016}, and Graczyk et al.~\cite{Graczyk2018} have demonstrated that dynamical coupling of a homogeneous ferromagnetic film with a  periodic array of ferromagnetic stripes placed underneath can lead to the formation of a magnonic band structure in the film. 
However, the effect of a resonator on the phase of the reflected wave has not yet been studied in magnonics; moreover, the conditions for the existence of Fabry-Perot resonances and their effect on both reflected and transmitted SWs remain almost unexplored \cite{Mieszczak2020}, while both may be key to creating a magnonic metasurface. 


In this work, we investigate theoretically the interaction of a narrow, subwavelength-width stripe placed above the edge of a homogeneously magnetized film  with propagating SWs and its influence on the phase shift of reflected SWs.
Using frequency-domain finite-element calculations and micromagnetic simulations we find that this shift depends on the width of the stripe in a non-trivial way: an overall slow and steady increase of the phase shift with stripe width is repeatedly interrupted by sharp phase jumps by 360$\degree$. Treating the stripe and the underlying film as a non-reciprocal waveguide supporting two pairs of counter-propagating modes, we formulate a semi-analytical model that explains this behavior as a consequence of Fabry-Perot resonances produced by one of these mode pairs. We also show that by varying the film-stripe separation it is possible to switch between resonances of different order. Our results point to the importance of Fabry-Perot resonances appearing in locally bilayered ferromagnetic elements, with potential applications for the control of SW propagation in magnonic devices.

The paper is organized as follows. 
In Sec.~\ref{sec:methods}, we present the geometry of the system under consideration and the numerical methods used in our study. In Secs.~\ref{sec:DispersionBilayers}--\ref{sec:reflection-from-edge}, we analyze the key physical effects occurring in individual components of the system of interest: SW dynamics in infinitely extended films and bilayers, SW resonances in finite-width stripes and SW reflection from edges of truncated films. In Secs.~\ref{sec:phase_vs_W}--\ref{sec:phase_vs_S}, which form the core of this work, we investigate SW reflection at the end of a film coupled to a finite-width stripe. We show how interactions between parts of the system discussed in the previous subsections make the phase shift of the reflected wave highly sensitive to stripe width and the stripe-film separation. Our results are summarized in Sec.~\ref{sec:conclusions}.

\section{Structure and methods\label{sec:methods}}
\subsection{Structure under consideration}

\begin{figure}[!t]
    \includegraphics[width=8.6cm]{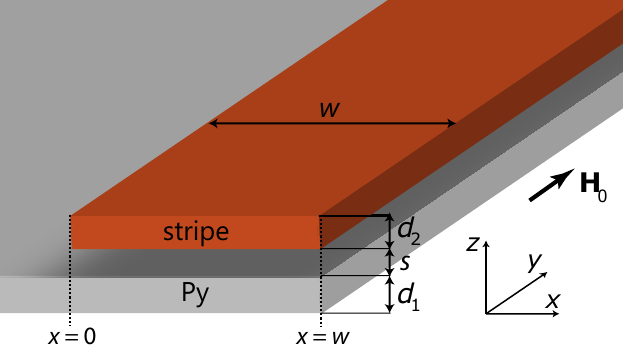}
    \caption{Geometry of the system used in simulations. A ferromagnetic stripe of width~$w$ and thickness~$d_2$ is separated from a semi-infinite permalloy film of thickness $d_1$ by a distance $s$. The right edges of both layers are aligned at $x=w$. The whole system is placed in vacuum in an external magnetic field $\mu_0 H_0= 0.1$ parallel to the $y$~axis. The geometry of the system is independent from~$y$.}
    \label{fig:geo_syst} 
\end{figure}

We consider a system composed of non-magnetic and ferromagnetic materials. Its geometry is independent of the $y$~coordinate and piecewise constant along $x$, as shown schematically in Fig.~\ref{fig:geo_syst}.
The system consists of a semi-infinite permalloy (Py) film of thickness 50~nm and a ferromagnetic stripe of thickness 40~nm and finite width~$w$. Both elements are separated by a distance $s$ and their right edges are aligned to $x=w$. Throughout the paper, we will vary the width~$w$ of the stripe and its separation~$s$ from the film. We are interested in manipulating the phase of the reflected SWs using subwavelength elements; therefore the width of the stripe will be smaller than or comparable to the wavelength of SWs in the Py film at the frequency of operation. The system is magnetized by a uniform in-plane bias magnetic field of magnitude $\mu_0 H_0= 0.1$~T directed along the $y$~axis.
In all calculations we have taken the saturation magnetization of the film to be $M\tsub S=760$~kA/m and its exchange constant, $A=13$~pJ/m. The stripe is made of a material (called FM2 from here on) with $M\tsub S=525$~kA/m and $A=30$~pJ/m;
lowering its saturation magnetization and increasing the exchange constant with respect to Py will allow us to exploit interactions of local resonances of the stripe with propagating waves in Py at frequencies characteristic for the dipolar and dipole-exchange SWs. Such a choice of the stripe's material shifts the SW spectrum down and flattens the first band for lower wavevectors with respect to the Py, as will be discussed in the next section.
The gyromagnetic ratio of both ferromagnets is $\gamma = -176$~rad GHz/T.

\subsection{Governing equations}

The magnetization dynamics is described by the Landau-Lifshitz equation:
\begin{equation}
    \partial_t \mathbf{M} = -\frac{|\gamma| \mu_{0}}{1+\alpha^2} \left[ \mathbf{M} \times \mathbf{H}_\mathrm{eff} 
    + \frac{\alpha}{M_\mathrm{S}} \mathbf{M} \times (\mathbf{M} \times \mathbf{H}_\mathrm{eff} )
    \right],
    \label{eq:LL_basic}
\end{equation}
where $\mathbf{M}$ is the magnetization vector, $\mu_0$ is the permeability of vacuum,  $\alpha$ is a dimensionless damping parameter, and 
\begin{equation}
    \mathbf{H}_\mathrm{eff}=\mathbf{H}_{0}+\mathbf{H}_\mathrm{m}+\mathbf{H}_\mathrm{ex}
    \label{eq:mag_field_com}
\end{equation}
is the effective magnetic field. The latter is the sum of the external magnetic field $\mathbf{H}_0$, the magnetostatic field $\mathbf{H}_\mathrm{m}$, and the isotropic Heisenberg exchange field  $\mathbf{H}_\mathrm{ex}=\boldsymbol\nabla \cdot (l^2\boldsymbol\nabla\mathbf{M})$, where $l=\sqrt{2A/(\mu_0M_\mathrm{S}^2)}$ is the exchange length. The magnetostatic field fulfils the magnetostatic Maxwell's equations 
\begin{subequations}
 \begin{align}
  \label{eq:gauss_law}
  \boldsymbol\nabla \cdot (\mathbf{M} + \mathbf{H}\tsub m) &= 0,\\
  \boldsymbol\nabla \times \mathbf{H}\tsub m &=0;
 \end{align}
\end{subequations}
the latter makes it possible to write it as $\mathbf H\tsub m = -\boldsymbol\nabla \varphi$, where $\varphi$~is the magnetic scalar potential.

Assuming a harmonic time dependence [$\exp(-\I \omega t$)], zero damping ($\alpha=0$) and alignment of the external magnetic field $H_0$ with the $y$~axis, splitting the magnetization $\mathbf M$ and magnetostatic field $\mathbf H\tsub m$ into static ($M\tsub S \hat{y}, H_0 \hat{y}$) and dynamic (radio-frequency) components ($\mathbf{m}=[m_x,0,m_z]$, $\mathbf{h}_{\mathrm{m}}=[-\partial_x \varphi,0,-\partial_z \varphi]$), linearizing the Landau-Lifshitz equation (\ref{eq:LL_basic}) (applicable only in the ferromagnetic layers) and coupling it with the Gauss law for magnetism, Eq.~(\ref{eq:gauss_law}) (applicable everywhere), we arrive at the following system of equations for the magnetic potential~$\varphi$ and the dynamic magnetization component~$\mathbf m$:
\begin{subequations}
\label{eq:linearised-equations}
\begin{align}
  \partial_x(m_x - \partial_x \varphi) + \partial_z(m_z - \partial_z \varphi) &= 0, \\  
  \partial_x \varphi - \vect\nabla \cdot (l^2 \vect\nabla m_x) + \frac{H_0}{M_S} m_x - \frac{\I \omega}{|\gamma| \mu_0 M_S} m_z &= 0, \\
  \partial_z \varphi - \vect\nabla \cdot (l^2 \vect\nabla m_z) + \frac{H_0}{M_S} m_z + \frac{\I \omega}{|\gamma| \mu_0 M_S} m_x &= 0.
\end{align}
\end{subequations}


\subsection{Numerical methods}

We have used three complementary numerical methods to study SW dynamics. First, we use micromagnetic simulations performed in the open-source environment mumax$^3$ \cite{vansteenkiste2014design}, which solves the full Landau-Lifshitz equation [Eq.~(\ref{eq:LL_basic})] with the finite-difference time-domain (FDTD) method. We use this method to calculate the dispersion relations of SWs and steady states obtainable after long continuous excitation of SWs by a specified source. 
More details can be found in Appendix~\ref{Sec:AppSimulations}. The main disadvantage of this method is its high computational cost. Resonant systems can take a long time to reach steady state, and the cost of a single time step is pushed up by the need to discretize the whole system on a uniform grid whose resolution is dictated by the size of the smallest geometric features.

In order to avoid these limitations, we rely primarily on calculations using the frequency-domain finite element method (FD-FEM). Its major advantage is the possibility of refining the mesh locally, e.g.\ only around small geometric features, rather than globally.  In addition, it allows direct and fast calculation of the eigenfrequencies and eigenmodes (mode profiles) of the system, which can be identified with its steady states. To perform the calculations, we have used the COMSOL Multiphysics software \cite{comsol} to solve the linearized Landau-Lifshitz equation coupled with the Gauss law, Eqs.~(\ref{eq:linearised-equations}), as described in \cite{rychly2017spin}. At the edges of the computational domain (far from the ferromagnetic materials) the Dirichlet's boundary conditions, forcing the magnetic potential to vanish, are imposed.

In Sec.~\ref{sec:model} we formulate a semi-analytical model dependent on the numerical values of scattering matrices associated with interfaces separating SW waveguides with different cross-sections. To calculate these matrices, we used the finite element modal method. In this method, Eq.~\eqref{eq:linearised-equations} is transformed into an eigenproblem whose solutions are the wavenumbers and profiles of propagative and evanescent modes of a particular $x$-invariant part of the system shown in Fig.~\ref{fig:geo_syst}, e.g.\ the Py film or the Py/FM2 bilayer. This eigenproblem is then discretized by expanding the magnetic potential and dynamic magnetization in a finite-element basis and solved numerically. Fields on each side of the interface are represented as superpositions of the eigenmodes of the respective parts calculated in the previous step. Finally, the scattering matrix is obtained by imposing the standard continuity conditions on these fields and solving the resulting system of equations for the excitation coefficients of outgoing modes.
This method, inspired by similar techniques used in photonics~\cite{liu1990}, will be described in  detail in a forthcoming paper~\cite{Smigaj}.

\section{Results}

\subsection{Dispersion relation of bilayers\label{sec:DispersionBilayers}}

\begin{figure}[t!]
    \includegraphics[width=8.6cm]{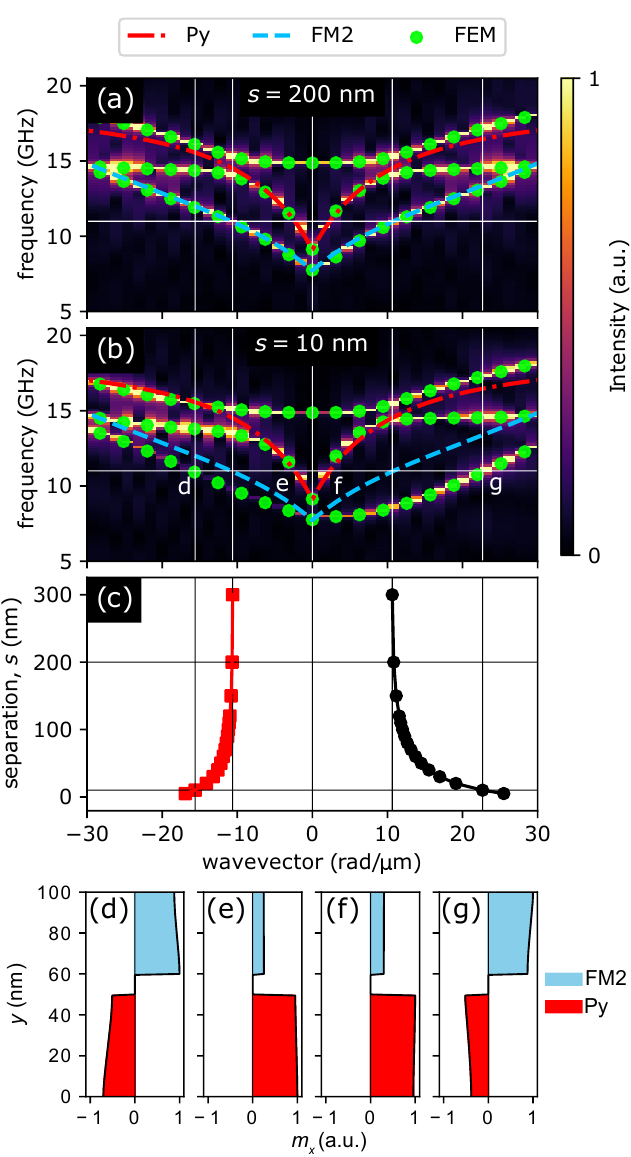}
    \caption{ (a)--(b) Dispersion relations calculated for two infinite films made of Py and FM2 separated by non-magnetic gap of width (a) 200 nm and (b) 10 nm. Colormaps in the background present results obtained by means of micromagnetic simulations, whereas green points correspond to the results of FD-FEM calculations. The dashed blue and the dash-dotted red lines are the analytical dispersion relations of SWs supported by isolated FM2 and Py films, respectively. The horizontal line marks the frequency $f=11$ GHz used in further calculations. (c) Dependence of the wavenumber of the slow modes, d and g, located predominantly in FM2, on the separation between films at frequency 11 GHz. (d)--(g) Mode profiles of $m_z$:  (d), (g) \textit{slow} and (e)--(f) \textit{fast} modes, in the Py/FM2 bilayer separated by a 10 nm non-magnetic gap at frequency 11 GHz  [marked in (b)].}
    \label{fig:disp}
\end{figure}

Before we study the coupling of a film with a finite-width stripe, let us first analyze the interaction between modes in infinitely extended films, i.e., in bilayers composed of an infinite Py film separated by a non-magnetic gap from another infinite film made of FM2.
Two example dispersion relations for gap widths $s=200$~nm and $s=10$~nm are presented in Fig.~\ref{fig:disp}(a) and~(b), respectively.  It is clear that for a 200~nm-wide gap, the SW modes in Py and stripe almost do not interact with each other: the calculated dispersion curve coincides with the analytical dispersion curves of isolated Py and FM2 films calculated using \cite[Appendix C.7]{stancil2009spin}: 
\begin{equation}
    \omega^2 = \omega_0  \left(  \omega_0 + \omega_\mathrm{M} \right) + \frac{\omega_\mathrm{M}^2}{4}\left[1 + \mathrm{e}^{-2kd} \right],
    \label{eq:dispersion}
\end{equation}
where $d$ is the film thickness, $\omega=2\pi f$ is the angular frequency of SWs ($f$ denotes the frequency),  $k$ is the wavenumber, $\omega_0=|\gamma| \mu_0 (H_0 + M_\mathrm{S}l^2 k^2)$ \cite[Chapter 7.1]{mag_osci_wav}, and $\omega_\mathrm{M}=|\gamma| \mu_0 M_\mathrm{S}$.

The only visible difference occurs at frequencies above 14~GHz 
where we can see  a hybridization between the fundamental SW mode and the first mode quantized across the thickness, a so-called perpendicular standing SW \cite{stancil2009spin}. This hybridization and perpendicular standing SWs, however, are not considered in the analytical model.

At frequencies below 14 GHz the modes in bilayered structure can be classified according to their origin and group velocity. The \textit{fast} modes are related to SW dynamics in Py [see Fig.~\ref{fig:disp}(e) and~(f)] and are characterized by steeper dispersion (therefore higher group velocity) and longer wavelengths. In contrast, the \textit{slow} modes originating in FM2 [see Fig.~\ref{fig:disp}(d) and~(g)] have lower group velocity and shorter wavelengths. 
It is worth noting that the wavelengths of SWs in separated  layers 
do not depend on the direction of propagation, while the dynamic dipolar coupling between SW modes in both layers combined with the nonreciprocal nature of surface SW modes introduces asymmetry \cite{mruczkiewicz2014nonreciprocal, gallardo2019, an2019optimization}. For $s=200$ nm the coupling is still very weak and at the frequency of 11~GHz that is used in further analysis and marked with the white horizontal line in Fig.~\ref{fig:disp}(a),   the fast and slow modes have wavelengths of 2660~nm and 590~nm, respectively, for both propagation directions.

Reduction of the non-magnetic gap width to $s = 10$~nm causes the modes of the individual films to interact much more strongly. The wavelength of the slow modes decreases significantly and their dispersion relation becomes strongly nonreciprocal; at 11 GHz, slow modes propagating leftwards have wavelength 390~nm and those propagating rightwards, 270~nm. Figure~\ref{fig:disp}(c) shows that the asymmetry of the dispersion relation for slow modes grows as the films are brought closer together. 

\subsection{Eigenmodes of finite-width stripes\label{sec:resonator}}
\begin{figure}[t!]
    \includegraphics[width=8.6cm]{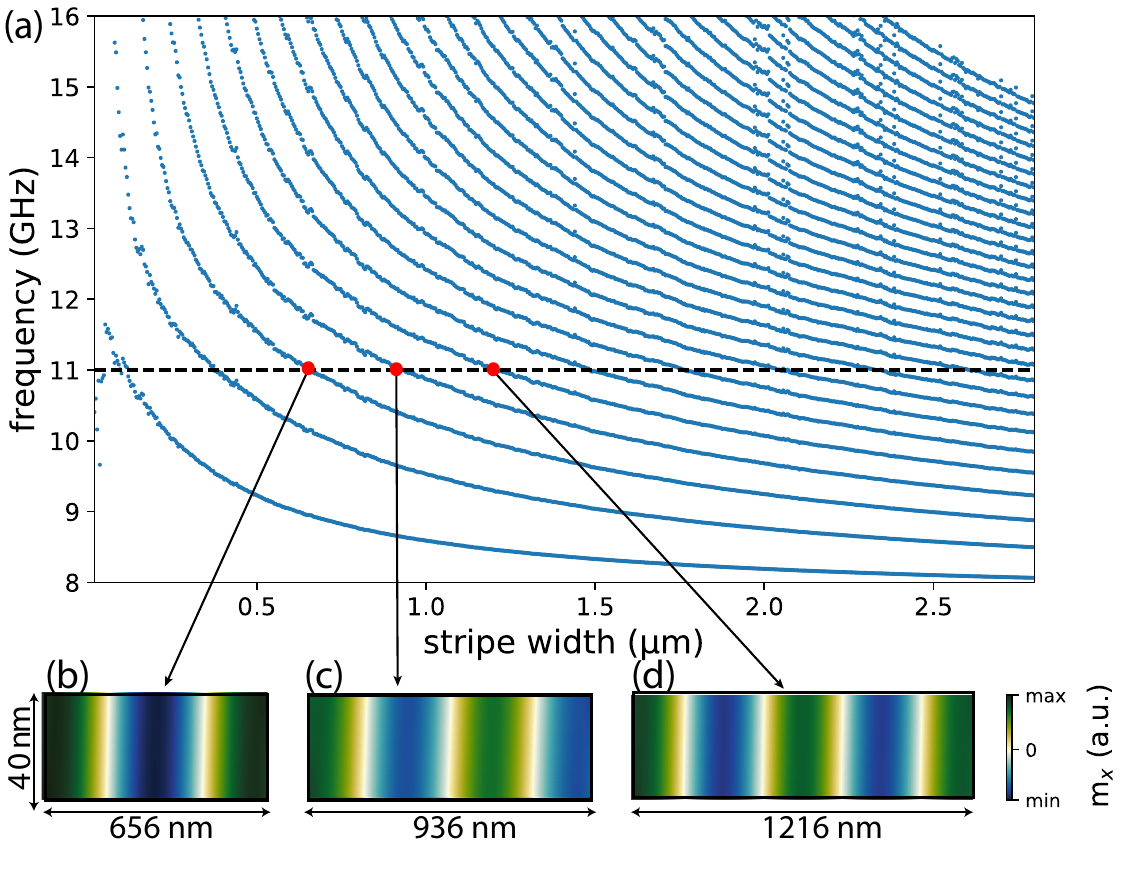}
    \caption{(a) Dependence of the frequency of stripe modes on stripe width. The dashed line marks the frequency $f=11$ GHz. (b), (c), and (d) Profiles of modes supported by stripes of width 656 nm, 936 nm, and 1216 nm at frequency 11 GHz. 
    }\label{fig:resonator-first-modes}
\end{figure}

The dependence of the eigenmode spectrum of isolated finite-width FM2 stripe on their width is displayed in  Fig.~\ref{fig:resonator-first-modes}(a). These calculations were made with FD-FEM for  the first thirty modes of stripes with widths up to 2800~nm (note that SW wavelength in uniform Py films at 11 GHz is 2660 nm). As intuitively expected, mode frequency decreases with increasing stripe width. 
The horizontal dashed line in Fig.~\ref{fig:resonator-first-modes} marks  $f=11$ GHz. It is visible that stripes of multiple widths support modes at this frequency. Profiles of three successive modes (for stripes of width 656~nm, 936~nm, and 1200~nm) are shown in Fig.~\ref{fig:resonator-first-modes}(b)--(d). Successive resonances appear for stripes of widths differing by approximately half of the wavelength of the eigenmode of a homogeneous  FM2 film, i.e., $\Delta w  \approx 280\;\mathrm{nm} \approx 0.5\lambda_{\mathrm{FM2}}$.

Lateral mode confinement in a reciprocal medium 
leads to formation of standing waves and quantization of the wavenumber. The standing waves have the form $\exp(\I k_n x) + \exp(-\I k_n x)$, where $n$ is the mode index, $k_n=r_n\pi/w$ and $r_n=n+\delta$ ($0\le\delta\le 1$).
For the Dirichlet (magnetic wall) boundary conditions, with the dynamic magnetization vanishing at the edges, we get $r_n=n+1$, whereas for ``free spins'' at the edges, $r_n=n$.  However, due to dipolar interactions, in magnetic stripes neither of these cases is correct and magnetization is partially pinned at the stripe edges, $0<\delta<1$ \cite{2002Guslienko}. This can also be interpreted as the effective length of the waveguide being slightly larger than the real one, or in terms of a non-zero phase shift~$\varphi$ being experienced at the stripe edges by the SWs forming the standing wave. Thus, resonances occur when the following condition is met:
\begin{equation}
    k w+\varphi = \pi n,\quad  n=1,2,\dots
    \label{eq:modeN_reciprocal}
\end{equation}
According to this equation, successive resonances at frequency 11~GHz should appear for stripes of widths differing by ca.\ $\lambda/2 \approx 280$ nm; this is confirmed by the FD-FEM calculations.

\begin{figure}[t!]
\includegraphics[width=8.6cm]{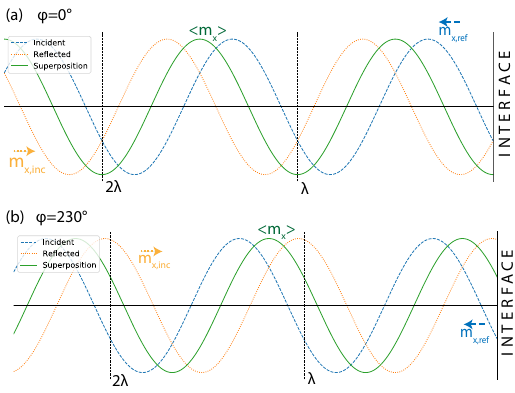}
\caption{ (a) Incident and reflected plane waves at an arbitrary time, and the intensity of the resulting interference pattern, in the zero phase shift case, i.e., $\varphi = 0^\circ$. The dotted orange line and the dashed blue line correspond to the incident and reflected plane waves, respectively. The solid green line corresponds to the SW intensity (time-averaged squared dynamic magnetization) of the interference pattern. 
(b) The same for a phase shift of $\varphi = 230^\circ$, i.e., the value obtained for SW reflection from the edge of a semi-infinite 50-nm-thick Py film.
}\label{fig:reflectionScheme}
\end{figure}

\subsection{Phase shift of the reflected SWs\label{sec:reflection-from-edge}}
Before studying the influence of the stripe's presence on the SW reflection, let us first discuss SW reflection from the edge of an isolated truncated film. According to Stigloher et al.~\cite{stigloher2018}, dynamic dipolar interactions induce a phase shift between the incident and reflected SWs. This phase shift is a natural consequence of the previously discussed dipolar pinning occurring at the boundaries of thin ferromagnetic film \cite{2002Guslienko}. Interestingly, Verba et al.~\cite{verba2020spin} have recently shown that a phase shift may also be introduced by a polarization mismatch between incident and reflected SW modes. Regardless of the physical mechanism responsible for the phase shift, we can extract its magnitude from steady-state solutions formed far away from the edge. 

The phase shift manifests itself in the resulting interference pattern as a displacement of nodes with respect to the interface from which the waves are reflected. If the interface is located at $x=x_0$ and the reflection coefficient is $\E^{\I\varphi}$, where $\varphi$~is the phase shift, the standing wave pattern sufficiently far from the interface (at $x \ll x_0$) will be
    \begin{equation}
      \begin{split}
        m(t; x) &= \RE \bigl\{ A \E^{-\I 2\pi f t}[\E^{\I k (x-x_0)} + \E^{\I\varphi}\E^{-\I k (x-x_0)} ] \bigr\} \\
        &= a(t) \cos[k(x-x_0)-\varphi/2],
        \label{eq:phaseShift}
      \end{split}
    \end{equation}
where $A$ and $a(t)$ are scaling coefficients independent of~$x$. The change in the standing wave pattern due to varying phase shift is illustrated in  Figs.\ \ref{fig:reflectionScheme}(a) and (b).

In practice, we calculate $\varphi$ by fitting the expression on the right-hand side of Eq.~(\ref{eq:phaseShift}) to a snapshot of $m_x$ on the symmetry axis of the Py film obtained from micromagnetic or FD-FEM simulations. To avoid distortions caused by evanescent waves excited at the interface, only points lying at least one stripe width 
to the left of the interface are taken into account. 

The phase shift occurring at the edge of a 50-nm-thick Py film at frequency 11 GHz is found numerically to be $230^\circ$. The resulting standing wave pattern is shown in Fig.~\ref{fig:reflectionScheme}(b). 

\subsection{Phase shift dependence on stripe width\label{sec:phase_vs_W}}

\begin{figure}[t!]
    \centering
    \includegraphics[width=8.6cm]{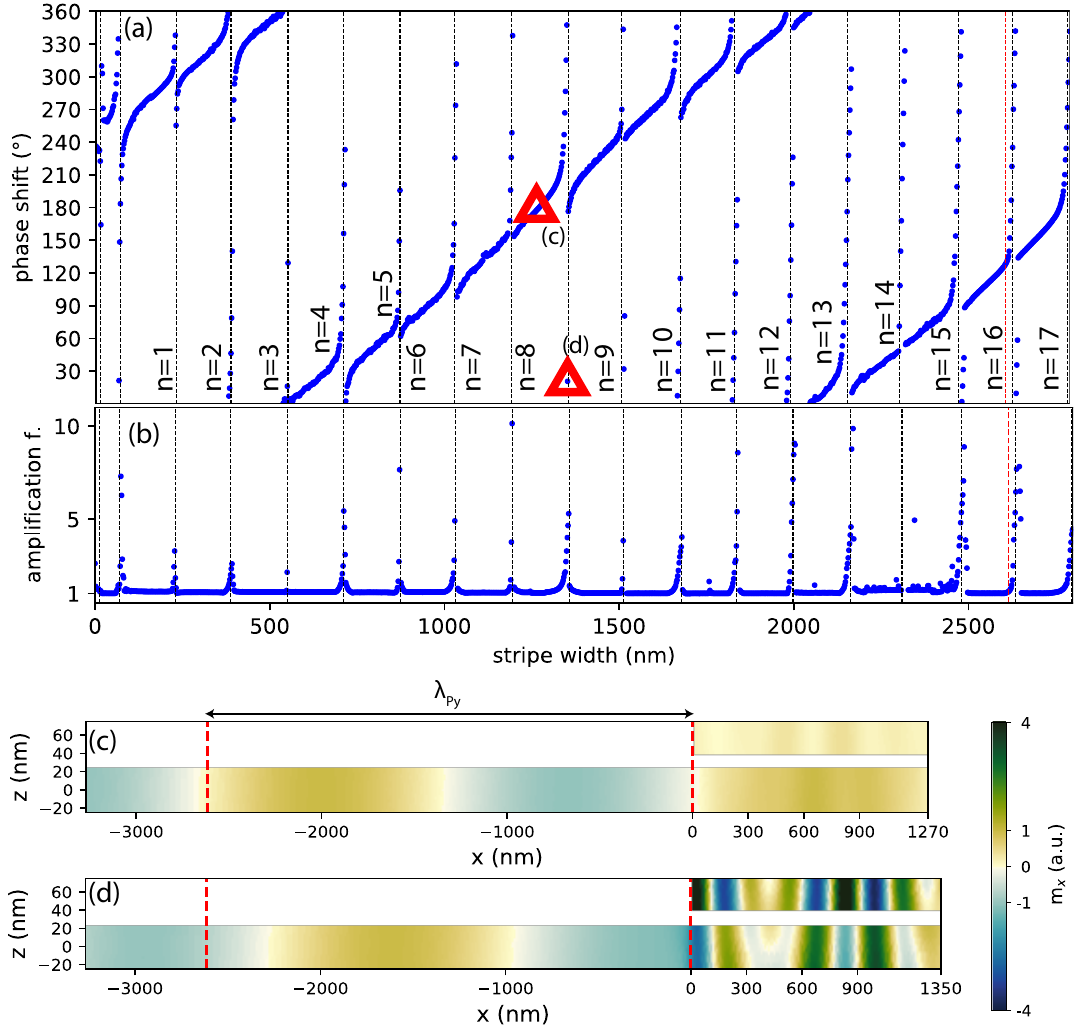}
    \caption{(a) Phase shift of SWs at frequency 11 GHz as a function of the stripe width, calculated by the FD-FEM. The stripe-film separation is $s=10$ nm. Resonances (rapid changes of the phase shift, marked with vertical dotted lines) appear periodically with a period of ca.~160~nm.
    (b) SW amplification factor in Py below the stripe, obtained by dividing the maximum of $|m_x|$ in the part of the film lying below the stripe by the maximum of $|m_x|$ in the far field, i.e., further than half of width of the Py layer to the left of the stripe. 
    (c) Snapshot of the dynamic magnetization $m_x$ in a plateau region (stripe width: 1270~nm). (d) Snapshot of the system in resonance at 11 GHz (stripe width: 1350~nm). The red dashed lines in (c) and (d) denote the SW wavelength in a uniform Py film at the same frequency. }
    \label{fig:com_res}
\end{figure}

The introduction of a stripe over the Py film edge locally modifies the environment in which SWs propagate due to dynamic dipolar interactions between the film and the stripe. In consequence, it influences also the phase shift of reflected waves. The variation of this phase shift with stripe width at frequency 11~GHz, calculated using FD-FEM, is plotted in Fig.~\ref{fig:com_res}(a). In general, this phase shift, defined according to Eq.~\eqref{eq:phaseShift} with interface defined at the left edge of the stripe, at $x_0 = 0$, grows steadily with stripe width, however with periodic  jumps by $360^\circ$. These jumps occur approximately every 160~nm and are accompanied by an increase of the amplitude of SWs in the Py film underneath the stripe, as shown in Fig.~\ref{fig:com_res}(b). In fact, at these stripe widths SWs are amplified in the whole bilayer, indicating that a resonant mode of the bilayer is excited. This can be seen by comparing snapshots of $m_x$ for stripes of width 1270~nm (slowly changing phase shift) and 1350~nm (rapidly changing phase shift)  presented in Figs.~\ref{fig:com_res}(c) and (d), respectively.

In addition to the enhancement of SW amplitude in the bilayer, Fig.~\ref{fig:com_res}(c) shows that the SWs in the stripe and in the underlying Py layer have approximately opposite phases. In view of the profiles of the \textit{fast} and \textit{slow} modes shown in Fig.~\ref{fig:disp}(d)--(g), this indicates that the slow modes dominate. 
The magnetization pattern in both layers is more complex than that of a typical standing wave composed of two counter-propagating waves with the same wavelength; indeed,  as discussed in Sec.~\ref{sec:DispersionBilayers}, the bilayer modes have an asymmetric dispersion relation\cite{zingsem2019}. For such a scenario, we can generalize the resonance condition~(\ref{eq:modeN_reciprocal}) 
to
\begin{equation}
    (k_{u} + k_{d})w + \varphi_{l}+\varphi_{r} = 2\pi n, \quad n=1,2,...,
    \label{eq:modeN_nonreciprocal}
\end{equation}
where $k_{u}$ and $k_{d}$ are the wavenumbers of right- and left-propagating modes, and  $\varphi_{l}$ and $\varphi_{r}$ are the phase shifts occurring at the left and right interfaces of the stripe. For $k_{u}=k_{d}=k$  and $\varphi_{r}=\varphi_{l}$ this equation reduces to Eq.~(\ref{eq:modeN_reciprocal}). Substituting here the wavelengths of the slow modes of a bilayer with $s=10$~nm given in Sec.~\ref{sec:DispersionBilayers}, we conclude that successive resonances should occur every 160 nm, which matches well the results of FD-FEM calculations shown in Fig.~\ref{fig:com_res}.

\begin{figure}[t!]
    \centering
    \includegraphics[width=8.6cm]{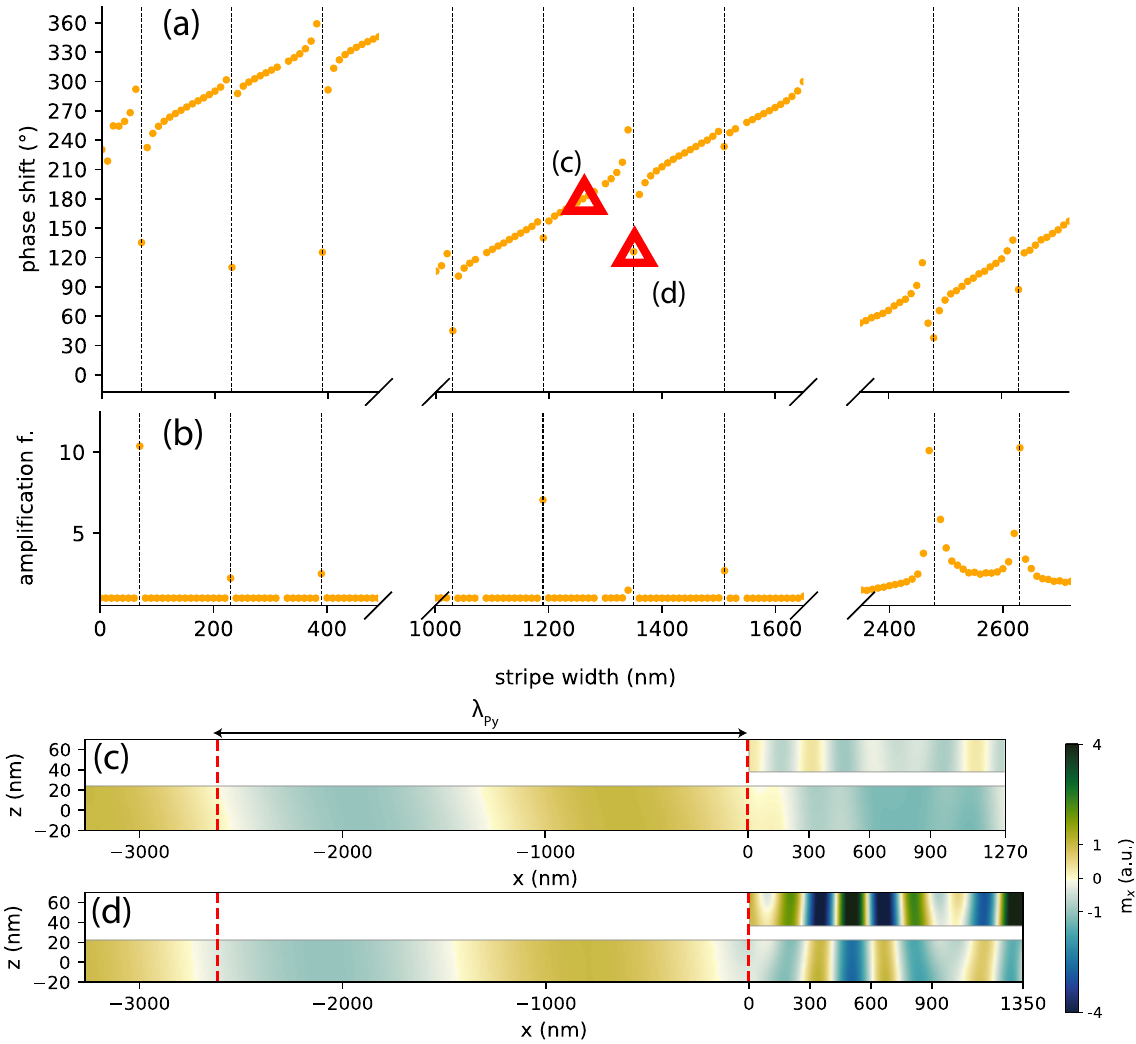}
    \caption{(a) Phase shift of SWs at frequency 11 GHz as a function of stripe width, obtained with micromagnetic simulations. (b) SW amplification factor in Py below the stripe, defined as in Fig.~\ref{fig:com_res}(b). (c) Snapshot of the dynamic magnetization $m_x$ in a system without a resonance at 11 GHz (stripe width: 1270~nm). (d) The same in a system with a resonance at 11 GHz (stripe width: 1350~nm). The red dashed lines in (c) and (d) denote the SW wavelength in a uniform Py film at the same frequency. 
    }
    \label{fig:mum_res}
\end{figure}

We have cross-checked these results against micromagnetic simulations made with a finite damping coefficient $\alpha=0.0001$, see Fig.~\ref{fig:mum_res}. Due to computational demands, these have been performed for a narrower range of stripe widths, 0--490 nm, 1000--1650 nm, and 2350--2720 nm, encompassing several resonances. The obtained results are consistent with those of FD-FEM calculations: the positions of resonances are the same as obtained by FD-FEM and the slope of the curve in intermediate regions is virtually identical. The almost perfect alignment of those results obtained by two different numerical methods confirms their correctness.


Snapshots of $m_x$ in systems with stripes of width 1270~nm and 1350~nm are displayed in Fig.~\ref{fig:mum_res}(c) and (d), respectively. The former does not have a resonance at the chosen frequency, whereas the latter does. The obtained magnetization patterns are qualitatively similar to those calculated by FD-FEM [Figs.~\ref{fig:com_res}(c) and (d)].

\subsection{Two-mode model analysis\label{sec:model}}

As discussed above, the plot of the phase shift vs.\ stripe width [Fig.~\ref{fig:com_res}(a)] shows areas of slow but steady growth separated by sharp resonances. The resonances are not identical: some, such as those at $n = 7$ and $n = 9$, are broader than others. These and other features can be explained by considering a semi-analytical model introduced below.

Wave scattering on the interface $x=0$ separating the film and the bilayer (see Fig.~\ref{fig:geo_syst}) can be described by a scattering matrix~$\mat S$ linking the complex amplitudes of the incoming and outgoing modes on both sides of the interface. If both parts are sufficiently long for the amplitudes of all incoming evanescent modes to be negligible, the amplitudes of the outgoing propagative modes are given by
\begin{equation}
 \label{eq:interface}
 \begin{bmatrix}
  d_1 \\ u_2 \\ u_3
 \end{bmatrix}
 = \mat S
 \begin{bmatrix}
  u_1 \\ d_2 \\ d_3
 \end{bmatrix}
 \equiv
 \begin{bmatrix}
  S_{11} & S_{12} & S_{13} \\
  S_{21} & S_{22} & S_{23} \\
  S_{31} & S_{32} & S_{33}
 \end{bmatrix}
 \begin{bmatrix}
  u_1 \\ d_2 \\ d_3
 \end{bmatrix}
 .
\end{equation}
Here, $u_1$ and $d_1$ are the amplitudes of the right- and left-propagating modes of the Py film, $u_2$ and $d_2$ are the amplitudes of the right- and left-propagating slow modes of the bilayer, and $u_3$ and $d_3$ are the amplitudes of the right- and left-propagating fast modes of the bilayer (see dispersion relation shown in Fig.~(\ref{fig:disp}). All these amplitudes are measured at the interface between the film and the bilayer. The elements of the scattering matrix~$\mat S$ can be calculated using the finite-element modal method~\cite{Smigaj}. At 11~GHz, their numerical values are 
\begin{equation}
 \mat S = 
 \begin{bmatrix}
   0.100-0.011\I & 0.135+0.016\I & 0.986+0.008\I \\
  -0.104-0.168\I & -0.290+0.929\I & 0.035-0.113\I \\
   0.975-0.023\I & 0.119+0.143\I & -0.118-0.0273\I
 \end{bmatrix}
\end{equation}
(these values are obtained for modes normalized to carry unit power, with the phase at the interface chosen so that $m_x$ is real and positive on the symmetry axis of the Py layer). It can be seen that the film mode is coupled primarily with the fast mode of the bilayer. The slow bilayer mode is strongly reflected. There is only weak, though non-negligible, coupling between the fast and slow bilayer modes. 

Likewise, the interface $x=w$ between the bilayer and the vacuum can be described by a scattering matrix $\mat S'$:
\begin{equation}
 \label{eq:interface'}
 \begin{bmatrix}
  d_2' \\ d_3'
 \end{bmatrix}
 = \mat S'
 \begin{bmatrix}
  u_2' \\ u_3'
 \end{bmatrix}
 \equiv
 \begin{bmatrix}
  S_{22}' & S_{23}' \\
  S_{32}' & S_{33}'
 \end{bmatrix}
 \begin{bmatrix}
  u_2' \\ u_3'
 \end{bmatrix}
 .
\end{equation}
Here, $u_2'$ and $d_2'$ are the amplitudes of the right- and left-propagating slow modes of the bilayer, and $u_3'$ and $d_3'$ are the amplitudes of the right- and left-propagating fast modes of the bilayer, all measured at the bilayer-vacuum interface (hence the prime, used to distinguish them from the amplitudes measured at the film-bilayer interface). The numerical values of these scattering coefficients calculated at 11~GHz are
\begin{equation}
 \mat S' = 
 \begin{bmatrix}
   -0.043+0.975\I & -0.103+0.190\I \\
   0.189+0.105\I & -0.561-0.799\I
 \end{bmatrix}
 .
\end{equation}
Both modes are strongly reflected and there is only weak cross-coupling.

Mode amplitudes at the two interfaces are linked by 
\begin{subequations}
 \label{eq:resonator}
 \begin{align}
  u_i' &= \exp(\I k_{iu} w)\, u_i \equiv \Phi_{iu} u_i,\\
  d_i &= \exp(-\I k_{id} w)\, d_i' \equiv \Phi_{id} d_i',\quad i = 2, 3,
 \end{align}
\end{subequations}
where $k_{iu}$ and $k_{id}$ are the wave numbers of the right- and left-propagating modes, numerically determined to be $k_{2u} = 16.2$, $k_{3u} = 2.22$, $k_{2d} = -23.1$ and $k_{3d} = -1.90$~rad/\textmu m.

Together, \eqref{eq:interface}, \eqref{eq:interface'} and \eqref{eq:resonator} form a system of nine equations for as many unknown mode amplitudes (the amplitude $u_1$ of the mode incident from the input film is treated as known). To obtain an intelligible expression for the reflection coefficient $r \equiv d_1 / u_1$, it is advantageous to start by eliminating the amplitudes $u_3$, $d_3$, $u_3'$ and $d_3'$ of the fast bilayer mode, which is only weakly reflected at the interface with the Py film and hence will not give rise to strong Fabry-Perot-like resonances. This mimics the approach taken by Lecamp \textit{et al.}~\cite{Lecamp} in their model of pillar microcavities. This reduces the second row of Eq.~\eqref{eq:interface} and the first row of Eq.~\eqref{eq:interface'} to
\begin{subequations}
\begin{align}
  u_2 &= \tilde S_{21} u_1 + \tilde S_{22} d_2, \\
  d_2' &= \tilde S_{22}' u_2' + \tilde S_{23}' \Phi_{3u} S_{31} u_1,
\end{align}
\end{subequations}
where
\begin{subequations}
 \label{eq:S-tilde}
 \begin{align}
  \begin{split}
    \begin{bmatrix}
    \tilde S_{21} & \tilde S_{22}
    \end{bmatrix}
    &\equiv 
    \frac{1}{1 - \kappa S_{23} \Phi_{3d} S_{32}' \Phi_{2u}}  \\
    &\quad\begin{bmatrix}
    S_{21} + \kappa S_{23} \Phi_{3d} S_{33}' \Phi_{3u} S_{31} &
    S_{22} + \kappa S_{23} \Phi_{3d} S_{33}' \Phi_{3u} S_{32}
    \end{bmatrix}
    ,
  \end{split}\\
  \begin{bmatrix}
   \tilde S_{22}' & \tilde S_{23}'
  \end{bmatrix}
  &\equiv 
  \frac{1}{1 - \kappa S_{23}' \Phi_{3u} S_{32} \Phi_{2d}}
  \begin{bmatrix}
   S_{22}' + \kappa S_{23}' \Phi_{3u} S_{33} \Phi_{3d} S_{32}' &
   \alpha S_{23}'   
  \end{bmatrix}
 \end{align}
\end{subequations}
and 
\begin{equation}
 \kappa \equiv (1 - S_{33} \Phi_{3d} S_{33}' \Phi_{3u})^{-1}.
\end{equation}
The fast bilayer mode is only weakly reflected at the interface with the film: $\abs{S_{33}} \approx 0.12 \ll 1$. Therefore multiple reflections of the fast mode at bilayer interfaces do not give rise to strong Fabry-Perot resonances and the coefficient $\kappa$ remains close to 1 for all bilayer lengths. Together with the fact that the cross-coupling coefficients $S_{23}$, $S_{32}$, $S_{23}'$ and $S_{32}'$ are small, this means we can expect the scattering coefficients with a tilde defined in Eq.~\eqref{eq:S-tilde} to be well approximated by
\begin{equation}
 \label{eq:S-tilde-approx}
 \begin{bmatrix}
  \tilde S_{21} & \tilde S_{22} \\
  \tilde S_{22}' & \tilde S_{23}'
 \end{bmatrix}
 \approx
 \begin{bmatrix}
  S_{21} + S_{23} \Phi_{3d} S_{33}' \Phi_{3u} S_{31} & S_{22} \\
  S_{22}' & S_{23}'
 \end{bmatrix}
 .
\end{equation}

Solving the equations remaining after elimination of the amplitudes of the fast mode for the amplitudes of the slow mode and substituting the resulting expressions to the formula for $d_1$ in the first row in Eq.~\eqref{eq:interface}, we arrive at the following formula for the reflection coefficient:
\begin{equation}
 \label{eq:r}
 r \equiv d_1/u_1 = (a + \beta b),
\end{equation}
where 
\begin{subequations}
 \label{eq:r-terms}
 \begin{align}
  \label{eq:a}
  a &\equiv \smallfactor{S_{11}} + S_{13} \Phi_{3d} \kappa S_{33}' \Phi_{3u} S_{31},\\
  \begin{split}
  \label{eq:b}
  b &\equiv \smallfactor{S_{12}} \Phi_{2d}(\tilde S_{22}' \Phi_{2u} \smallfactor{\tilde S_{21}} +
  \smallfactor{\tilde S_{23}'} \Phi_{3u} S_{31}) \\
  &\quad + S_{13} \Phi_{3d} \kappa \bigl[
    S_{33}' \Phi_{3u} \smallfactor{S_{32}} \Phi_{2d} 
    (\tilde S_{22}' \Phi_{2u} \smallfactor{\tilde S_{21}} + \smallfactor{\tilde S_{23}'} \Phi_{3u} S_{31}) \\
    &\quad + \smallfactor{S_{32}'} \Phi_{2u} 
    (\smallfactor{\tilde S_{21}} + \tilde S_{22} \Phi_{2d} \smallfactor{\tilde S_{23}'} \Phi_{3u} S_{31})\bigr]
  \end{split}
 \end{align}
\end{subequations}
and $\beta$ represents the effect of multiple reflections of the slow mode:
\begin{equation}
 \label{eq:beta}
 \beta \equiv (1 - S_{22} \Phi_{2d} S_{22}' \Phi_{2u})^{-1}.
\end{equation}
To facilitate the interpretation of Eqs.\ \eqref{eq:r}--\eqref{eq:beta}, the scattering coefficients with magnitude much smaller than~1 have been underlined.

\begin{figure}
    \centering
    \includegraphics[width=8.6cm]{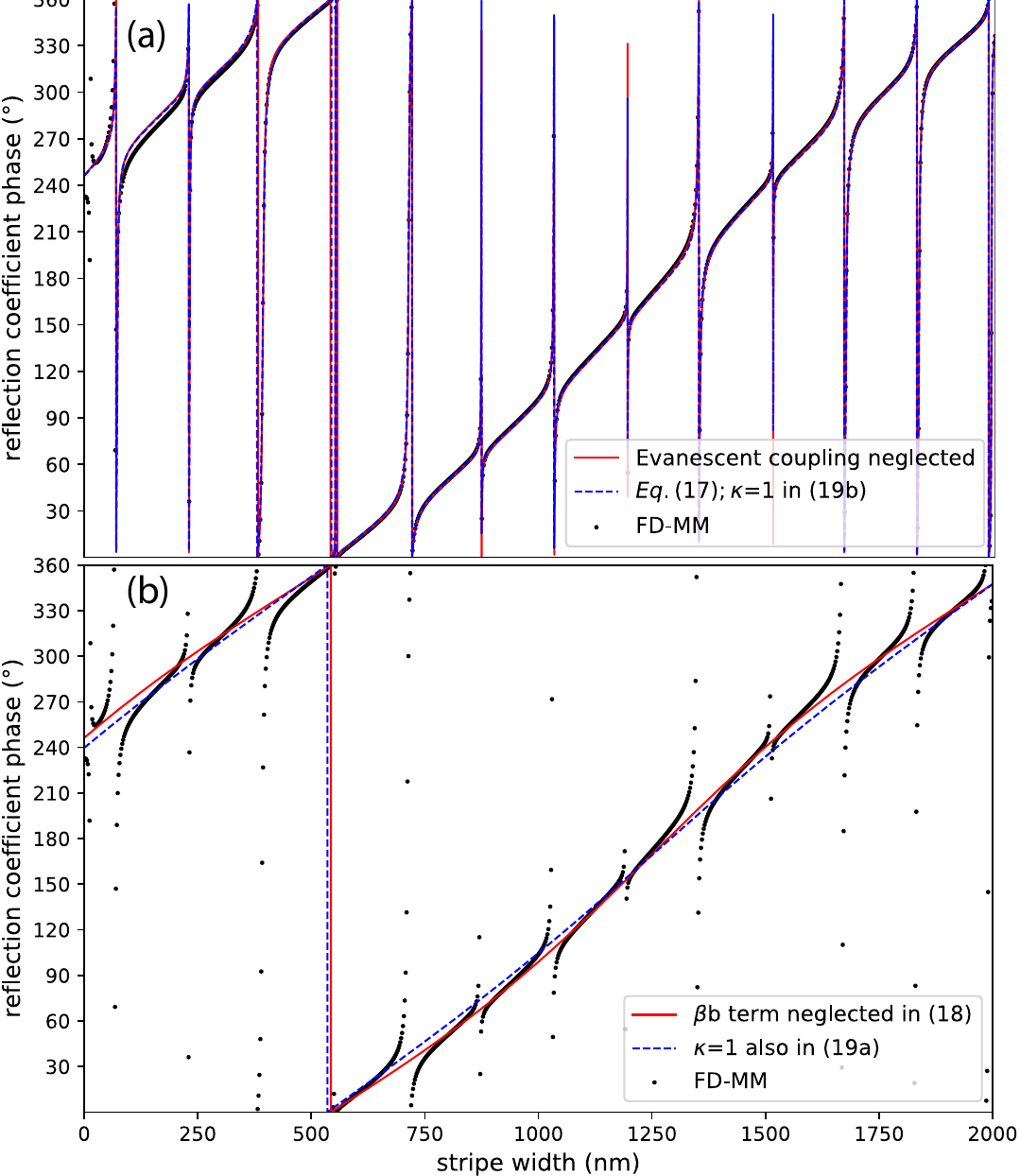}
    \caption{\label{fig:approximate-r}Comparison of the reflection coefficient phase calculated numerically using the finite-element modal method FD-MM (black points) with the semi-analytical model from Eq.~\eqref{eq:r} at varying degrees of approximation (color lines on both subplots). Details in the plot's legends and in the text. 
    }
    \label{fig:fdmm}
\end{figure}

It can be seen that the reflection coefficient~$r$ is made up of two terms. The first, $a$, is dominated by the phase shift acquired by the fast mode of the bilayer during a single round-trip across it. This term produces the slow but steady increase of the phase shift visible in Fig.~\ref{fig:com_res}(a) (also in Fig.~\ref{fig:approximate-r}). The second term, $\beta b$, is proportional to $b$, which is a superposition of six small terms, each containing a product of two scattering coefficients of small magnitude. Therefore $\beta b$ has an appreciable effect on the reflection coefficient~$b$ only when the factor~$\beta$, representing the combined effect of multiple reflections of the slow mode on both ends of the bilayer, is much greater than 1. This happens at stripe widths~$w$ corresponding to Fabry-Perot resonances of the slow mode, where $[\arg S_{22} + \arg S_{22}' + (k_{2u} + k_{2d})w]$ is a multiple of $2\pi$, justifying the postulated resonance condition Eq.~\eqref{eq:modeN_nonreciprocal}. Since $b$ is a combination of multiple terms of similar magnitude, its dependence on the stripe width is rather complicated. This explains the variability of the shapes of individual resonances in Fig.~\ref{fig:com_res}(a) [also in Fig.~\ref{fig:approximate-r}(a)]. 

To confirm this interpretation of the role of the various terms in Eq.~\eqref{eq:r}, let us visualize and compare the effects of applying successively stronger approximations to it. In Fig.~\ref{fig:approximate-r}, the black symbols show the variation of the phase of the reflection coefficient obtained directly from numerical calculations made with the finite-element modal method [in close agreement with the FD-FEM results from Fig.~\ref{fig:com_res}(a)]. The red solid curve in \ref{fig:fdmm}(a) shows the phase of the reflection coefficient calculated from Eq.~\eqref{eq:r}. The only approximation made in its derivation was to neglect evanescent coupling between the left and right end of the bilayer; clearly, this approximation is very well satisfied everywhere except for stripes narrower than 250~nm. The blue dashed curve in \ref{fig:fdmm}(a) shows the effect of applying the approximation~\eqref{eq:S-tilde-approx} and setting $\kappa$ to 1 in the formula \eqref{eq:b} for~$b$ (but not in the formula \eqref{eq:a} for~$a$). This corresponds to neglecting terms proportional to products of more than two small scattering coefficients; the resulting curve is almost indistinguishable from the previous one. Neglecting the second term $\beta b$ in Eq.~\eqref{eq:r} produces the red solid curve in \ref{fig:fdmm}(b). The resonances are gone, but the long-term increase in phase shift with stripe width is still reproduced faithfully. Finally, the blue dashed curve in \ref{fig:fdmm}(b) shows the result of approximating $\kappa$ by 1 also in the formula \eqref{eq:a} for~$a$. Its small deviation from the red curve confirms the minor role played by multiple reflections of the fast mode.

\begin{figure}[t!]
    \centering
    \includegraphics[width=8.6cm]{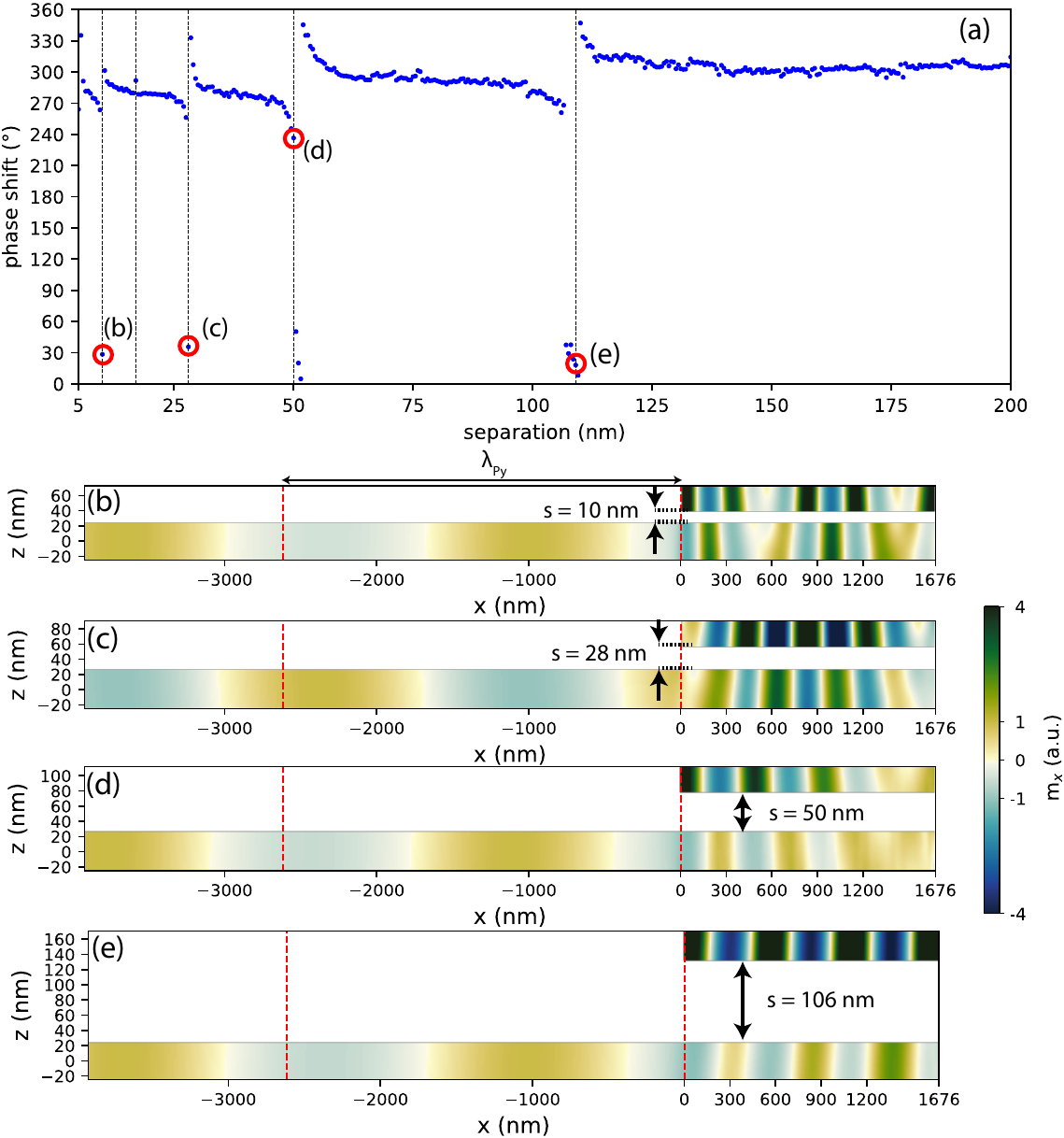}
    \caption{(a) Phase shift dependence on the separation between the FM2 stripe and Py film, calculated by the FD-FEM. 
    (b)--(e) Snapshots of the dynamic magnetization $m_x$ at separations $10$~nm, $28$~nm, $50$~nm and $106$~nm, corresponding to the resonances marked in plot (a). The dashed red lines mark the SW wavelength in the Py film.
    }
    \label{fig:altitude}
\end{figure}

\subsection{Phase shift dependence on the layer separation\label{sec:phase_vs_S}}
In Sec.~\ref{sec:DispersionBilayers} we observed that the separation influences the strength of the dynamical dipolar coupling between the infinitely wide stripe and Py film. It affects the SW dispersion relation, especially the wavelengths of the slow modes propagating leftwards and rightwards. Therefore, according to Eq.~(\ref{eq:modeN_nonreciprocal}), by varying the separation $s$, and hence $k_u$ and $k_d$, while keeping the stripe width~$w$ constant, it should be possible to sweep over resonances of different orders~$n$. Indeed, we have found multiple resonances in  dependence of the phase shift on separation~$s$ for a stripe of width $w= 1676$ nm, as shown in Fig.~\ref{fig:altitude}.

Resonances do not appear periodically; the spacing between subsequent Fabry-Perot resonances increases with the altitude of the stripe, and for the chosen stripe width the last resonance occurs at the separation $s=106$~nm. This is because with increasing $s$ the coupling between the Py film and the stripe weakens and the wavenumbers $k_u$ and $k_d$ approach their asymptotic limits. Snapshots of the magnetization at the resonances found at separations $10$~nm, $28$~nm, $50$~nm and $106$~nm are shown in Figs.~\ref{fig:altitude}(c)--(e). These figures demonstrate a clear enhancement of the SW amplitude below the stripe and a decrease in the number of nodal point with increasing separation, in line with the shift of the dispersion relation of the slow mode towards smaller wavenumbers shown in Fig.~\ref{fig:disp}(c).

    \begin{figure}[t!]
        \centering
        \includegraphics[width=7cm]{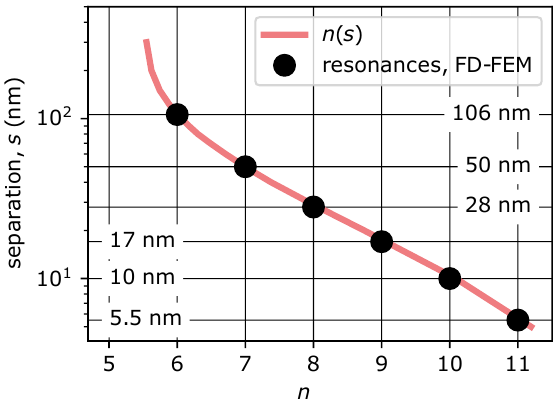}
        \caption{Red line: dependence of the resonance index $n$ calculated from Eq.~\eqref{eq:modeN_nonreciprocal} on the separation~$s$, with $\varphi_{\mathrm{l}}+\varphi_{\mathrm{r}}$ 
        set to the fitted value $0.12$. Resonances are predicted to occur at integer values of~$n$.
        Black circles: positions of resonances found in FD-FEM calculations shown in Fig.~\ref{fig:altitude}. }
        \label{fig:modeNumber_onSeparation}
    \end{figure}

Combining the Fabry-Perot resonance condition, Eq.~(\ref{eq:modeN_nonreciprocal}), with the numerically calculated dispersion relations and setting $\varphi_\mathrm{l}+\varphi_\mathrm{r}$ 
to the fitted value of  $0.12$, 
we calculate the dependence of $n$ 
on the separation between the stripe and the Py film, plotted with the red line in  Fig.~\ref{fig:modeNumber_onSeparation}. In the range from 106 down to 5 nm, we find six integer values of $n$, corresponding to  successive resonances. These values agree well with the results of FD-FEM calculations, where six resonances, marked with black points in Fig.~\ref{fig:modeNumber_onSeparation}, are detected in that range of separation.


\section{Conclusions\label{sec:conclusions}}


We have studied theoretically the influence of a narrow ferromagnetic stripe of subwavelength width placed at the edge of a ferromagnetic film on the phase shift of reflected SWs. 
At the considered frequency (11 GHz) the bilayer formed by the film and the stripe supports two pairs of slow (short wavelength) and fast (long wavelength) guided SW modes propagating in opposite directions; these modes couple with the SW mode of the Py film. This allowed us to interpret the numerical results by modelling the system as a series of waveguides linked by junctions at which waveguide modes are scattered into each other.

We have found a strong nonlinear dependence of the phase shift on the stripe width. We showed that the reflection coefficient, from which the phase shift can be derived, consists of two terms, each having a different origin. One produces a slow but steady increase of the phase shift with increasing bilayer width and is dominated by the phase accumulated by the fast mode  during a single round-trip across the bilayer. The other term has an appreciable effect on the reflection coefficient only when multiple reflections of the slow mode  on both edges of the bilayer interfere constructively, which corresponds to Fabry-Perot resonances of this mode. Interestingly, the incoming wave from the Py film couples strongly only to the fast mode, but at Fabry-Perot resonances, the phase of the reflected SW is controlled by the weakly coupled slow mode. 
Essentially, this system is a realization of a Gires–Tournois interferometer \cite{Gires1964} operating on SWs. However, in this design, its width is smaller than the wavelengths of the incident waves and the interferometer utilitizes two nonreciprocal SW modes present in the bilayer.

We have also found that the phase shift of the reflected SW passes through a series of resonances as the separation between the stripe and the Py film is increased. This unexpected effect originates from the dependence of the wavelength of the slow SW in the bilayer on the strength of the dipolar coupling between the two layers. As a result, the bilayer width at which the resonance Fabry-Perot condition is satisfied changes with the coupling strength as well, giving rise to the separation-dependent resonances.  

Overall, this research shows that SW Gires–Tournois interferometer can be used to modify the phase of reflected SWs in a wide range by tiny changes of the bilayer part width or stripe-film distance. This is significant for the further development of magnonic devices where SW phase control is of key importance, in particular in  integrated systems with components smaller than the SW wavelength. This may include the use of arrays of resonators in tunable SW optical elements, such as lenses,  magnonic metasurfaces and phase shifters, as well as the sensing applications of magnonics, for example the development of magnonic counterparts of sensors utilizing surface plasmon resonances.

\subsection{Acknowledgments}
The research leading to these results has received funding
from the Polish National Science Centre projects No. 
UMO-2015/17/B/ST3/00118,
UMO-2019/33/B/ST5/02013, and  UMO-2019/35/D/ST3/03729. The simulations were partially performed at the Poznan Supercomputing and Networking
Center (Grant No. 398).

\appendix
\section{Micromagnetic simulations}\label{Sec:AppSimulations}

Micromagnetic simulations were performed in the mumax$^3$ environment \cite{vansteenkiste2014design} for the same magnetic parameters and geometry as described in Sec.~\ref{sec:methods} and damping $\alpha=0.0001$. The simulated structure was discretized on a mesh consisting of regular $5\times 100 \times 5$ nm$^3$ unit cells. 
In order to model a film infinitely extended along the $y$ axis,
we have imposed periodic boundary conditions along the $y$ axis with assumed 1024 repetitions of the system along the $y$ axis. 

After stabilizing the system with a magnetic field of value 0.1~T applied along the $y$~axis, SWs were excited in the film by a local source of microwave-frequency magnetic field of frequency 11~GHz and amplitude 0.1~mT placed at 3.6~\textmu m from the right edge. 
To prevent wave reflections from the left edge of the film, an absorbing zone with gradually increasing damping was defined on the left side of the system. 
A continuous harmonic SW excitation was maintained for 162.6~ns in order to reach a fully evolved (steady-state) interference pattern of the incident and reflected waves.


\bibliography{literature}

\end{document}